\newcommand{\diracslash}[1]{#1\llap{/\kern2pt}}

\newcommand{\be}{\begin{equation}}
	\newcommand{\ee}{\end{equation}}
\newcommand{\bea}{\begin{eqnarray}}
	\newcommand{\eea}{\end{eqnarray}}
\newcommand{\ba}[1]{\begin{array}{#1}}
	\newcommand{\ea}{\end{array}}

\documentclass[prd,11pt,aps,floats,nofootinbib,tightenlines,showpacs]{revtex4-1}
\usepackage{epsfig,graphicx,pstricks}
\usepackage{psfrag}
\usepackage{color}
\usepackage{amsmath}
\usepackage{amsfonts}
\usepackage{amssymb}
\usepackage{textcomp}
\usepackage{multirow}
\usepackage{subfigure}
\usepackage[breaklinks=true,colorlinks=true,linkcolor=blue,urlcolor=blue,citecolor=blue]{hyperref}
\usepackage[normalem]{ulem}

\newcommand{\mcomment}[1]{}

\begin{document}

%\thispagestyle{plain}

%\markboth{WSPC}{Using World Scientific's style file}

%\section{Usage}

%\begin{verbatim}

	\title {Conserved charge susceptibilities in the relativistic mean-field hadron resonance gas model: constraints on hadronic repulsive interactions}
		\author{Somenath Pal}
		\address{Variable Energy Cyclotron Centre, 1/AF, Bidhan Nagar , Kolkata-700064, India}
%	\email{}
	
	\author{Guruprasad Kadam}
	\address{Department of Physics and Material Science and Engineering,  Jaypee Institute of Information Technology, A-10, Sector-62, Noida, UP-201307, India}

	\author{Abhijit Bhattacharyya}
	\address{Department of Physics, University of Calcutta, 92, A.P.C. Road, Kolkata-700009, India}
%	\email{}
	
\def\be{\begin{equation}}
\def\ee{\end{equation}}
\def\bearr{\begin{eqnarray}}
\def\eearr{\end{eqnarray}}

	\begin{abstract}
 We investigate the effect of repulsive interaction between hadrons on the susceptibilities of conserved charges, namely baryon  number (B), electric charge (Q) and strangeness (S). We estimate  second  fourth and sixth order susceptibilities of conserved charges,  their differences, ratios and correlations within ambit of mean-field hadron resonance gas (MFHRG) model.  We consider repulsive mean-field interaction among meson pairs, baryon pairs and anti-baryon pairs separately and constrain them  by confronting the results of various susceptibilities with the recent lattice QCD (LQCD) data. We find that the repulsive interactions between baryon-baryon pairs and antibaryon-antibaryon  pairs are sufficient to describe the baryon susceptibilities of hadronic matter at temperatures below QCD transition temperature.  However,  small but finite mesonic repulsive interaction is needed  to describe electric charge and strangeness susceptibilities. We finally conclude that the repulsive interaction between hadrons play a very important role in describing the thermodynamic properties of hadronic matter, especially near the quark-hadron phase transition temperature ($T_c$). The mean-field parameter for baryons  ($K_B$) should be constrained  in the range $0.40\le K_B\le 0.450$ $\text{GeV.fm}^{3}$ to get a good agreement of baryon susceptibilities with the LQCD results, whereas meson mean-field parameter $K_M\sim 0.05$ $\text{GeV.fm}^{3}$ must be included with $K_B$ to get a reasonable agreement of MFHRG model with the LQCD results for electric charge and strangeness susceptibilities.

	\end{abstract}

\maketitle

\section{Introduction}
 Currently accepted gauge theory of strongly interacting matter is quantum chromodynamics (QCD), in which, the  fundamental constituents  are  (colored) quarks and gluons. If we have a macroscopic system composed of quarks and gluons then its  thermodynamic state is specified by four external control parameters, namely temperature (T), and three chemical potentials ($\mu_B, \mu_Q, \mu_S$)  corresponding to the conservation of baryon number (B), electric charge (Q) and strangeness quantum number (S). In a plot of temperature ($T$) against baryon chemical potential ($\mu_B$) - which is known as QCD phase diagram - every point  corresponds to an equilibrium state. QCD phase diagram is largely a conjectured one and understanding it, theoretically as well as experimentally,   is one of the main objectives of elementary particle physics research today\cite{Ding:2015ona,Bzdak:2019pkr,Ratti:2021ubw}.

The heavy-ion collision experiments (HIC), namely Large Hadron Collider (LHC), the Relativistic Heavy-Ion Collider (RHIC) etc., have been very successful in probing  a part of the QCD phase diagram, where the QCD coupling constant is large. In these experiments, the thermodynamic state of QCD matter as it existed at the time of hadronization, is probed. The thermodynamic parameters specific to hadronization  stage in the evolution can be extracted either from the particle yields or from higher order cumulants characterizing the distribution of produced hadrons at freeze-out. These observables are compared with  the statistical thermal model calculations, namely hadron resonance gas model (HRG). In this model, the equilibrium thermodynamic state of low temperature and low density hadronic matter can be effectively described by an ideal gas of point-like hadrons and resonances\cite{Dashen:1969ep, Dashen:1974jw, Welke:1990za, Venugopalan:1992hy}. HRG model has been tremendously successful in describing low temperature hadronic phase of QCD\cite{braun2004quark}. Some studies have discussed the effect of magnetic field on hadronic equation of state within the ambit of HRG model\cite{Bhattacharyya:2015pra,Kadam:2019rzo,Pradhan:2021vtp}. It has successfully described  the hadron yields as measured in HIC experiments\cite{Braun-Munzinger:1994ewq, Braun-Munzinger:1995uec, Braun-Munzinger:1999hun,Cleymans:1996cd,Cleymans:1999st,Becattini:2005xt,Braun-Munzinger:2001hwo,Andronic:2005yp,Andronic:2008gu} and 
reproduced the equation of state (EoS) calculated in lattice QCD (LQCD) simulations at zero as well at small but finite $\mu_B$\cite{Karsch:2003vd,Karsch:2003zq,Tawfik:2004sw,Huovinen:2009yb,Alba:2015iva,Huovinen:2017ogf,Vovchenko:2014pka}. 

Fluctuations and correlations of conserved charges, namely baryon number (B), electric charge (Q) and strangeness (S) have been sensitive probes of deconfinement~\cite{Bazavov:2013uja}. There have been several studies in this regard~\cite{Bhattacharyya:2013oya,Bhattacharyya:2017gwt,Bhattacharyya:2014uxa,Fukushima:2008wg,Roessner:2006xn,Sasaki:2006ww,Ratti:2007jf,Fukushima:2009dx,Schaefer:2006ds,Schaefer:2009ui,Schaefer:2007pw,Wambach:2009ee,Motornenko:2020yme,Pal:2020ink,Pal:2020ucy,ab1,ab2,beta,Pal2024,PhysRevC.96.025205,RAU2014176,PhysRevD.97.114030}. They can also be used to calculate thermodynamic quantities at finite baryon density via Taylor series expansion\cite{Petreczky:2012rq,Ding:2015ona}. Despite its success recent studies show that the differences between actual QCD calculations and that of ideal HRG model become appreciable in the properties of higher order susceptibilities. At higher temperatures, the particle densities are higher and virial expansion only up to second order coefficient may not be sufficient to describe the thermodynamic properties. The validity of HRG model at higher temperatures can be justified by detailed comparison with the recent LQCD results. The fluctuations and correlations of conserved charges have been studied in the LQCD\cite{Borsanyi:2010bp,Borsanyi:2011sw,HotQCD:2012fhj,Bazavov:2013dta,Bazavov:2013uja,Bazavov:2014xya,Bazavov:2014yba, Borsanyi:2014ewa, Bellwied:2015lba, Ding:2015fca, Bazavov:2015zja, DElia:2016jqh,Bazavov:2017dus,Borsanyi:2018grb} and results from HRG model have been used as a benchmark. It is found that the higher order susceptibilities estimated within ambit of ideal HRG model shows significant deviations from LQCD results near $T_c$. The prominent reason for this discrepancy is the exclusion of short-range repulsive interactions between hadrons as shown in the recent studies which include repulsive interaction in the ideal HRG model~\cite{Karthein:2021cmb}. These repulsive interactions are especially important at higher densities and make appreciable changes in the EoS as described by hadron resonance gas model. Various approaches to account for the short range repulsive interactions have been proposed, namely excluded-volume approach\cite{Rischke:1992rk}, van der Waals interaction approach~\cite{Vovchenko:2020lju}, relativistic mean-field approach\cite{Kapusta:1982qd,Olive:1980dy}, relativistic virial expansion~\cite{Huovinen:2017ogf} etc. All these different methods, to incorporate the repulsive interactions on the HRG model, can be shown to be different approximations of a more general phenomenological framework~\cite{Anchishkin:2014hfa}.

The parameters which specify the repulsive interactions in a given model can be constrained by confronting the model results with the LQCD calculations. Higher order susceptibilities related to conserved charges, namely baryon number (B), electric charge (Q) and strangeness (S) are the important thermodynamic quantities to constrain the QCD effective models. Further, these susceptibilities are related to the cumulants. Recently, such studies have been carried out to constrain the excluded volume parameter ($b$) of excluded-volume HRG (EVHRG) model \cite{Bollweg:2021vqf}. The repulsive interactions can be constrained by constraining the excluded volume parameter $b$. It was found that  $b>1\: \text{fm}^3$ gives a reasonable agreement of  EVHRG results with the LQCD results at vanishing chemical potential for second order susceptibilities. In the present work, we calculate the conserved charge susceptibilities (up to order six), their ratios, differences and correlations  within ambit of relativistic mean-field HRG model (MFHRG). In this model, repulsive interactions are accounted via density dependent mean-fields. This gives a shift in the single particle energy, $U=K n$, where $n$ is the number density and $K$ is the parameter determining the strength of the repulsive interaction. We shall discuss the validity of MFHRG model by comparing it with the recent LQCD results for the second order susceptibilities. This analysis, as we shall see,  constrains the mean-field parameter $K$ which is related to the mean-field potential for the repulsive interaction between hadrons. With the best fit values of  $K$ we shall confront the other higher order susceptibilities of MFHRG model with the available lattice results. We shall consider baryon-baryon (antibaryon-antibaryon) repulsive interactions parameterized via mean-field parameter $K_B$ and meson-meson repulsive interaction  via mean-field parameter $K_M$. We shall neglect the repulsion between baryon and mesons in our analysis.

This paper is organized as follows: In section \ref{sec2}, we present a brief description of  hadron resonance gas model with repulsive mean-field interactions and calculation of various susceptibilities. In section  \ref{sec3}  we shall discuss the results, and finally in section  \ref{sec4} we summarize our findings.
\section{ Mean-field Hadron Resonance Gas (MFHRG) model}
\label{sec2}
Ideal HRG model  takes into account only the attractive interactions between hadrons by including the heavy resonance states in the partition function. Repulsive interactions between the hadrons can be incorporated in this model  via mean-field approach. In this approach  the single particle energies are shifted by a term proportional to the particle number density as
\begin{equation}
	\varepsilon_{a}=\sqrt{p^2+m_{a}^2}+U(n)=E_{a}+U(n)
	\label{dispersion}
\end{equation}
where $n$ is the total hadron number density. The potential energy $U$ represents repulsive interaction between hadrons. We assume that the potential energy is proportional to the number density $n$ as\cite{Kapusta:1982qd}
\begin{equation}
	U(n)=Kn
	\label{poten}
\end{equation}
Here, $K$ is a constant parameter which characterize the strength of repulsive interaction. Although, one can explore different power law expressions for $U(n)$, we take the simplest form which is effective when the hadron gas is not very dense~\cite{Savchuk:2020yxc}. We assume different repulsive interaction parameters for baryons $(K_B)$ and mesons $(K_M)$ such that
\begin{equation}
	U(n_{B\{\bar{B}\}})=K_Bn_{B\{\bar{B}\}} \hspace{1cm} ( \text{baryons(antibaryons)})
	\label{potenbar}
\end{equation}

\begin{equation}
	U(n_M)=K_Mn_M  \hspace{1cm} (\text{mesons})
	\label{potenmes}
\end{equation}
Where $n_M$ and $n_B$ are total number densities for mesons and baryons respectively. For baryons,
\begin{equation}
	n_{{B}}(T,\mu_B,\mu_{Q},\mu_S)=\sum_{a\in B}\int d\Gamma_{a}\:\frac{1}{e^{\frac{(E_{a}-{\mu_{\text{eff},B}})}{T}}+1}
	\label{numdenbaryon}
\end{equation}
where $\mu_{\text{eff},{B}}=c_i\mu_i-K_Bn_{B}$ and $c_i =(B_i,Q_i,S_i), {\mu_i}=(\mu_B,\mu_Q ,\mu_S)$ and $d\Gamma_{a}\equiv\frac{g_{a}d^{3}p}{(2\pi)^3}$, $g_a$ is the degeneracy of $a^{\text{th}}$ hadronic species. Similarly, the number density of antibaryons is

\begin{equation}
	n_{{\bar{B}}}(T,\mu_B,\mu_{Q},\mu_S)=\sum_{a\in \bar{B}}\int d\Gamma_{a}\:\frac{1}{e^{\frac{(E_{a}-{\mu_{\text{eff},\bar{B}}})}{T}}+1}
	\label{numdenantibaryon}
\end{equation}
where $\mu_{\text{eff},{\bar B}}=\bar{c_i}{\mu_i}-K_Bn_{\bar{B}}$. Note that we further assume same repulsive mean-field parameter for baryons and anti-baryons. For mesons,

\begin{equation}
	n_{{M}}(T)=\sum_{a\in M} \int d\Gamma_{a}\:\frac{1}{e^{\frac{(E_{a}-{\mu_{\text{eff},M}})}{T}}-1}
	\label{numdenmeson}
\end{equation}
where $\mu_{\text{eff},{M}}={c_i}\mu_i-K_Mn_{M}$.  Eqs. (\ref{numdenbaryon})-(\ref{numdenmeson})  are self consistent equations and can be numerically solved. The contributions to the pressure from baryons and mesons are then given, respectively, as
\begin{eqnarray}
	P_{B\{\bar{B}\}}(T,\mu_B,\mu_{Q},\mu_S)&=&T\sum \limits_{a\in B \{\bar B \}} \int d\Gamma_{a}
	\text{ln}\bigg[1+\text{Exp}\bigg(-\frac{(E_a-\mu_{\text{eff}}\{\bar{\mu}_{\text{eff}}\})}{T}\bigg)\bigg]
	\nonumber \\&-&\phi_{B\{\bar{B}\}}(n_{B\{\bar{B}\}})
	\label{premf}
	\label{pbbar}
\end{eqnarray}
\begin{eqnarray}
	P_M(T)&=&-T \sum_{a\in M} \int d\Gamma_{a}\text{ln}\bigg[1- {e^{\frac{(E_{a}-{\mu_{\text{eff},M}})}{T}}}\bigg] \nonumber \\
	&-&\phi_M(n_M)
	\label{nummf}
\end{eqnarray}

where,
\begin{equation}
	\phi_B(n_{B\{\bar{B}\}})=-\frac{1}{2}K_Bn_{B\{\bar{B}\}}^2
\end{equation}
and
\begin{equation}
	\phi_M(n_M)=-\frac{1}{2}K_Mn_M^2
\end{equation}

It is now useful to obtain an approximate expressions for the thermodynamic quantities. For that purpose we expand the logarithm in the expression for (non-interacting HRG) pressure in powers of  fugacity $z=\exp ({\beta\mu_{eff}})$ so that the baryon pressure in  Eq.(\ref{pbbar}) can be written as
\bearr
\frac{{P}_{B\{\bar B\}}}{T^4}&=&\sum\limits_{a\in B\{\bar B\}}\frac{g_a}{2\pi^2}(\beta m)^2\sum\limits_{l=1}^{\infty}
(-1)^{l+1} \:l^{-2} \times \mathcal{K}_2(\beta l m_a)z^{l} \nonumber\\
&+&\frac{K_BT^2}{2}\left(\frac{n_{B\{\bar B\}}}{T^3}\right)^2
\label{bessell}
\eearr
where $ \mathcal{K}_2$ is the Bessel function.  It can be easily  shown that as long
as $\beta(m_a-\mu_{eff})\gtrsim 1$, the contribution to the pressure $P^{id}_{B\{\bar B\}}$
 can be approximated by the leading term i.e. $l=1$ in the summation which, in fact, corresponds to Boltzmann approxiation. In this limit, the pressure from the baryons become
\bearr
&\frac{P_{B\{\bar B\}}}{T^4}
=\sum_{a\in B}\frac{g_a}{2\pi^2}(\beta m_a)^2K_2(\beta m_a)\nonumber\\
&\times \exp(\beta\mu_{eff}^a)+\frac{K_BT^2}{2}\left(\frac{n_{B\{\bar B\}}}{T^3}\right)^2
\eearr

The number density for baryons, Eq.(\ref{numdenantibaryon}),    can be written as   (again, as long as $\beta(m_a-\mu_{eff})\gtrsim 1$ i.e the Boltzmann approximation), 
\be
\frac{n_{B}}{T^3}=\sum_{a\in{B}}\frac{g_a}{2\pi^2}(\beta m)^2K_2(\beta m_a)e^{\beta\mu_{eff}^a}
\label{approxnb}
\ee

The baryon (antibaryon) pressure (Eq.(\ref{bessell})) can now be written as

\begin{equation}
P_{B\{\bar B\}}=Tn_{B\{\bar B\}}+\frac{K_B}{2}n_{B\{\bar B\}}^2
\label{pre_appx2}
\end{equation}

For a typical phenomenological values of mean-field parameter $K_B$, the quantity $\beta U $ turns out to be very small in the temperature range in which we are interested. Therefore, we can expand the exponential in the equations for $n_{B\{\bar B\}}$ to first order in $K$. Thus, in leading order of $K_B$ the baryon number density can be written as

\begin{equation}
n_{B\{\bar B\}}=n_{B\{\bar B\}}^{id}\bigg(1-\beta K_B n_{B\{\bar B\}}^{id}\bigg)
\end{equation}

Here, $n_B^{id}$ is the ideal gas number density and is given by
\be
\frac{n_{B}^{id}}{T^3}=\sum_{a\in{B}}\frac{g_a}{2\pi^2}(\beta m_a)^2K_2(\beta m_a)\exp(\beta c_a^i\mu_i)
\label{approxnbid}
\ee
where the sum is over all the baryons and resonances.  The total pressure due to baryons and antibaryons pressure (Eq. \ref{pre_appx2}), to leading order in $K_B$, can  be written as
\begin{equation}
P^{\text{tot}}_{(B,\bar{B})}(T,\mu_B,\mu_{Q},\mu_S)=T(n_B^{\text{id}}+ n^{\text{id}}_{\bar B})-\frac{K_B}{2}\bigg[(n_B^{\text{id}})^2+ (n^{\text{id}}_{\bar B})^2\bigg]
\label{preappx}
\end{equation}
 It may be noted that the effect of the density dependent repulsive  interaction essentially  reduces the pressure as compared to the ideal gas at finite densities. The (scaled) total pressure from baryons and antibaryons can then be written in more compact form as

\begin{eqnarray}
&&\frac{P_B+P_{\bar B}}{T^4}= \sum_{a\in B}F_a(\beta m_a)\cosh(\beta c_a^i\mu_i)\nonumber\\
&-&\frac{K_B}{2}\left(\sum_aG_a(\beta m_a,\beta\mu_Q,\beta\mu_s)\right)e^{2\beta\mu_B}\nonumber\\
&-&\frac{K_B}{2}\sum_a\left(G_a(\beta m_a,-\beta\mu_Q,-\beta\mu_s)\right)e^{-2\beta\mu_B}.
\label{pbapprox}
\end{eqnarray}

Here we have defined the  chemical potential independent function $F_a(\beta m_a)$ as
\be
F_a(\beta m_a)=\frac{g_a}{\pi^2}(\beta m_a)^2 K_2(\beta m_a)
\label{Fa}
\ee
and, the baryon chemical potential independent function
\bearr
&&G_a(\beta m_a,\beta\mu_Q,\beta\mu_s)=\frac{g_a}{2\pi^2}(\beta m_a)^2 K_2(\beta m_a)\nonumber\\
&\times &\exp(Q_a\beta\mu_Q+S_a\beta\mu_s)
\label{Ga}
\eearr

\section{Results and Discussion}
\label{sec3}

  Conserved charge susceptibilities are defined as
\begin{equation}
\chi^{mn}_{ij} = \frac{\partial^{m+n} (P(T,\mu_i,\mu_j)/T^4)}{\partial \left({\mu_i/T}\right)^m \partial \left({\mu_j/T}\right)^n} 
\end{equation}
where $i,j=B,Q,S$ corresponding to baryon number, electric charge and strangeness respectively. In the leading order approximation of the pressure (Eq.\ref{preappx}), the baryon number susceptibilities, baryon-strangeness correlation and baryon-electric charge correlations are respectively given by\cite{Huovinen:2017ogf} 

\begin{equation}
\chi_{B}^n=(\chi_{B}^{\text{id}})^n-2^n\beta^4(n_B^{\text{id}})^2 \hspace{1cm} (\text{even}\: n)
\end{equation}

\begin{equation}
\chi_{BS}^{n1}=(\chi_{BS}^{\text{id}})^n+2^{n+1} \beta^5 K_B n_B^{\text{id}}\sum_{j}B_j S_j P_{j}^{\text{id}} \hspace{1cm} (\text{odd}\: n)
\end{equation}

\begin{equation}
\chi_{BQ}^{n1}=(\chi_{BQ}^{\text{id}})^n+2^{n+1} \beta^5 K_B n_B^{\text{id}}\sum_{j}B_j Q_j P_{j}^{\text{id}} \hspace{1cm} (\text{odd}\: n)
\end{equation}
where, $P_{j}^{\text{id}}=Tn_{j}^{\text{id}}$ is the partial pressure of $j^{\text{th}}$  species.  
In the above equations, the first term represents purely non-interacting part and the second term represents contribution from the repulsive interaction.

Recently,  different sets of hadron lists compiled from PDG 2016 and 2020 tables have been used to confront the thermodynamics of HRG model with LQCD results. In the present work, we shall use a latest version of the list, namely  QMHRG2020 used in reference~\cite{Bollweg:2021vqf}. This list includes all the confirmed (3-and 4-star baryon resonances) and unconfirmed (1- and 2-star baryon resonances as well as mesons not listed in the PDG 2020 summary tables) hadrons. Furthermore, QMHRG2020 list also includes  quark-model states in the strange and non-strange baryon sectors.
%\begin{enumerate}
%\item PDG2016/ PDG2020: List of only well known states (**** and ***) from  Particle Data Booklet.
%\item PDG2016+: List of well established and unconfirmed states (** and *).
%\item Quark model: List of hadrons predicted by the quark model.
%\item Quark model HRG (QMHRG): List of well known hadrons (2016 and 2020 list) which is augmented with a list of QM states in the strange and non-strange baryons.
%\end{enumerate}

MFHRG model is characterized by two parameters, namely $K_M$ and $K_B$. We first confront the MFHRG results of second order susceptibilities with the recent LQCD results to establish the best fit values of these parameters. We adopt a simple version of the MFHRG model in which mean-field interactions are included only for meson-meson pairs, baryon-baryon pairs and antibaryon-antibaryon pairs. We neglect repulsive interaction between baryons and mesons. This corresponds to  minimal mean-field extension of the HRG model. Note that the meson-baryon interactions are presumed to be dominated by resonance formation, which is  the basis of ideal HRG model by construction. For the baryon–antibaryon system the short-range repulsive interactions are unlikely due to annihilation processes, which are included in the hadron resonance gas at equilibrium . The absence of short-range repulsion in the baryon–antibaryon system leads to only a small correction, since mesons dominate at small $\mu_B$ and since there are very few antibaryons at large $\mu_B$\cite{Andronic:2012ut}.

\begin{figure}[h]
\vspace{-0.4cm}
\begin{center}
%\begin{tabular}{c c c}
 \includegraphics[scale=0.45]{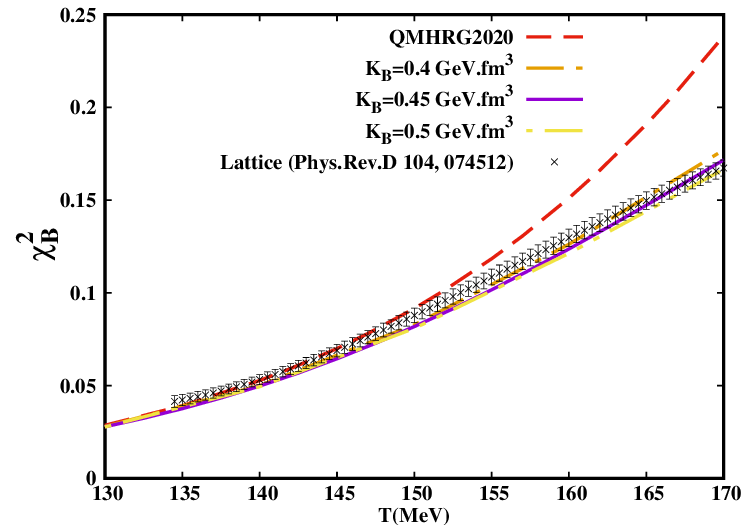}
  \includegraphics[scale=0.45]{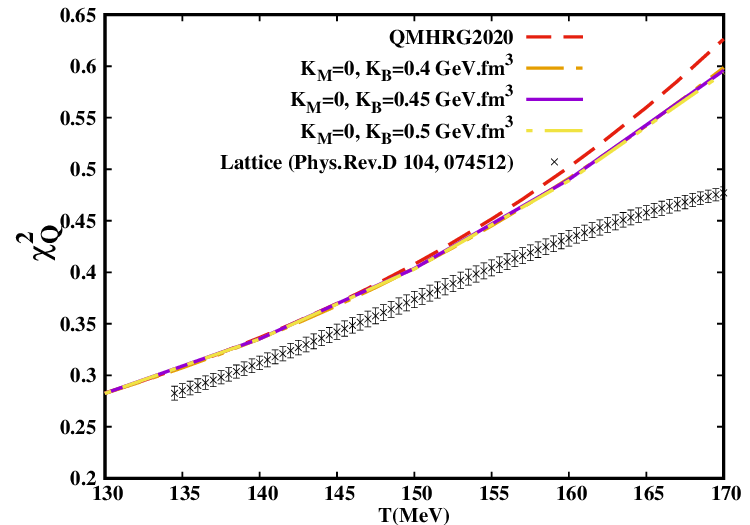} \\
    \includegraphics[scale=0.45]{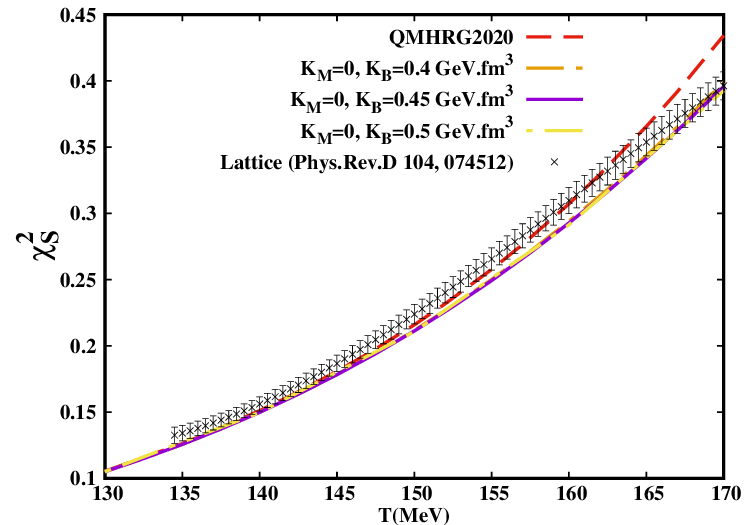}
%  \end{tabular}
	\caption{ Temperature dependence of second order susceptibilities of conserved charges (at $\mu_B=\mu_Q=\mu_S=0$) for different values of baryon mean-field parameter $K_B$ and for vanishing meson mean-field parameter $K_M$. Lattice data is  taken from Ref.~\cite{Bollweg:2021vqf}. QMHRG2020 model corresponds to particle list  compiled from PDG2020 list augmented with quark model states in the strange and non-strange baryons\cite{Bollweg:2021vqf}. Note that the mesons do not contribute to the baryon susceptibilities.}
		\label{chi2kb}.
  \end{center}
 \end{figure}

In the hydrodynamic description of matter produced in heavy-ion collision, the equation of state of mean-field HRG has been discussed in reference \cite{Sollfrank:1996hd}. Values discussed in this reference, $K=0.450\:\text{GeV.fm}^3$ and $0.660\:\text{GeV.fm}^3$ correspond to $T_c=165$ MeV and 140 MeV respectively. In the current work, we have compared the results of the HRG model with the lattice QCD results with $T_c\sim 155$ MeV. So, it is reasonable to restrict $K_B = 0.4 - 0.5\text{GeV.fm}^3$. Past studies show that the strength of mesonic repulsive interaction is small compared to that of baryons\cite{Vovchenko:2016rkn}. The validity of choice of $K_M$, however, can only be judged {\it{a posteriori}}..

Figure \ref{chi2kb} shows the results of second order susceptibilities estimated within HRG and MFHRG model and compared with LQCD results\cite{Bollweg:2021vqf}. Dashed red curve corresponds to HRG model which, when compared with LQCD data, overestimates $\chi_B^2$ after $T\sim 150$ MeV.  Initially we  fix $K_M=0$ and vary $K_B$ from 0.4 to 0.5 $\text{GeV.fm}^{3}$  in the steps of 0.05. We note that as we increase the value of $K_B$,  $\chi_B^2$  is suppressed at high temperature as compared to ideal HRG model. The best fit is observed for $K_B=0.45$  $\text{GeV.fm}^{3}$ (purple curve). However, with $K_M=0$, MFHRG overestimates   $\chi_Q^2$. As we will see later in this section,  we need a small but non-zero value of the meson mean-field parameter to get a better agreement with the lattice data. In case of $\chi_S^2$, ideal HRG results are in better agreement with the LQCD results up to $T\sim 160$ MeV.  MFHRG model slightly underestimates $\chi_S^2$, however, reproduces the general behavior observed in LQCD. We might get better agreement with of MFHRG with the lattice data if we choose different $K_B$ for strange and non-strange mesons. EVHRG model with different hard-core radii of strange and non-strange hadrons have been recently discussed in reference\cite{Motornenko:2020yme}. However, this variation in MFHRG will require additional fitting parameter which will make it more complicated and spoils the simplicity of the model.

\begin{figure}[h]
\vspace{-0.4cm}
\hspace{-0.8cm}
\begin{center}
\begin{tabular}{c c}
 \includegraphics[scale=0.5]{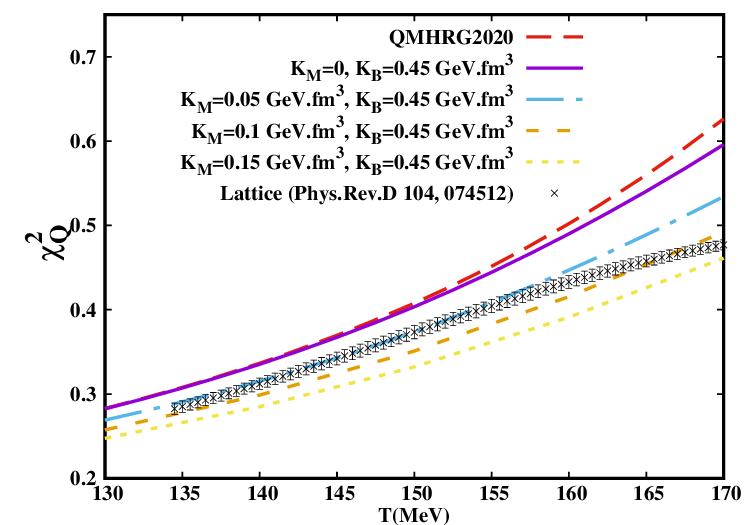}&
  \includegraphics[scale=0.5]{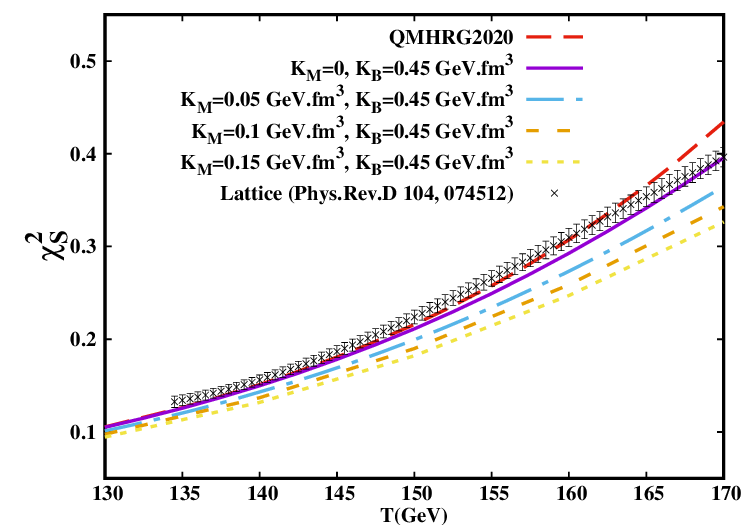}
  \end{tabular}
 \caption{ Temperature dependence of $\chi_Q^2$ and $\chi_S^2$ (at $\mu_Q=\mu_S=0$) for different values of meson mean-field parameter $K_M$ and for a fixed value of baryon mean-field parameter $K_B$. Lattice data is  taken from Ref.~\cite{Bollweg:2021vqf}. Note that we have excluded  $\chi_B^2$ because  variation in  $K_M$ does not affect it.}
		\label{chi2km}
  \end{center}
 \end{figure}

Figure \ref{chi2km} shows the results of second order susceptibilities, $\chi_Q^2$ and $\chi_S^2$, with  $K_B=0.45$ $\text{GeV.fm}^{3}$ and  for different values of $K_M$ ranging from 0 to 0.15 $\text{GeV.fm}^{3}$ in the steps of 0.05. Note that mesons do not contribute to $\chi_B^2$ and hence the variation in $K_M$ won't have any effect on this susceptibility. As noted earlier, both HRG model and MFHRG model with $K_M=0$ overestimate $\chi_Q^2$ over all the temperature range under consideration. However, if we assign small but non-zero value of meson mean-field parameter $K_M=0.05$ $\text{GeV.fm}^{3}$, we get good agreement with the LQCD data up to $T\sim 160$ MeV. If we increase $K_M$ further, even slightly,  MFHRG underestimates the LQCD data due to stronger suppression from the repulsive interactions. In case of $\chi^2_S$, HRG model results already being in good agreement with LQCD, assignment of small mean-field value to mesons, strongly suppress it.

\begin{figure}[h]
\vspace{-0.4cm}
\hspace{-0.8cm}
\begin{center}
\begin{tabular}{c c}
 \includegraphics[scale=0.5]{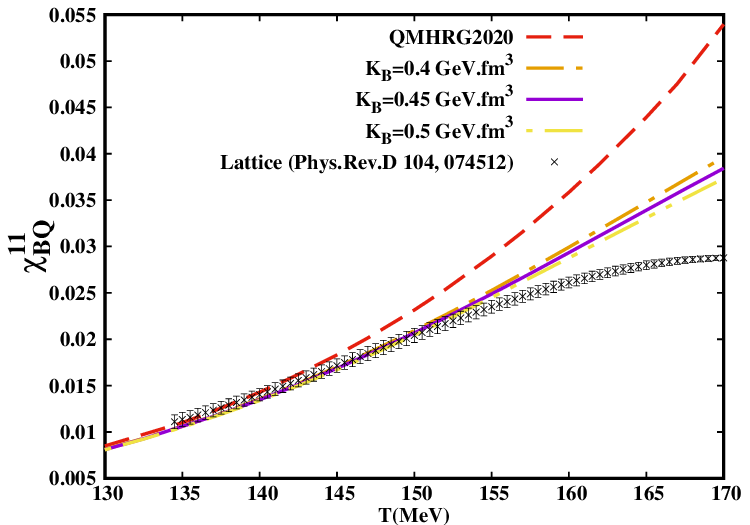}&
  \includegraphics[scale=0.5]{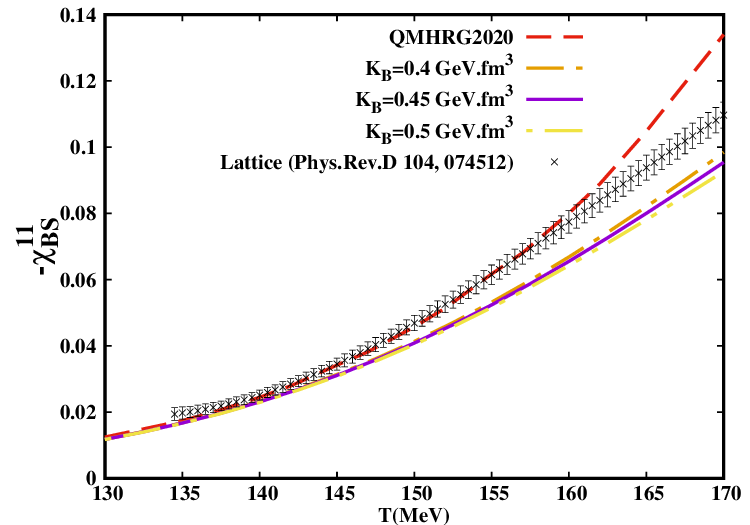}
  \end{tabular}
\caption{Temperature dependence of baryon-charge correlation $\chi_{BQ}^{11}$ and baryon-strangeness correlation $\chi_{BS}^{11}$. Lattice data is taken from \cite{Bollweg:2021vqf}.}
		\label{bsq}
  \end{center}
 \end{figure}

Fig.\ref{bsq} shows BQ correlation $\chi_{BQ}^{11}$ and BS correlation (-)$\chi_{BS}^{11}$. We note that ideal HRG model overestimates BQ  correlations, but if we include repulsive interactions, we get reasonable agreement with the LQCD data. These repulsive interactions accounted via mean-fields suppress the thermodynamic quantities as compared to ideal HRG model. At low temperatures,  suppression is relatively small, and hence, there is no appreciable differences between ideal HRG and MFHRG. The appreciable suppression occurs only at high temperatures when sufficient number of heavy-baryons are produced.   However, in case of $\chi_{BQ}^{11}$, inclusion of repulsive interaction in the ideal HRG model improves the agreement with LQCD results only up to $T\sim 150$ MeV.  Further, in case of $\chi_{BS}^{11}$, ideal HRG results are in good agreement with the LQCD results up to $T\sim 150$ MeV, whereas,  MFHRG model  results deviate from LQCD results  above $T\sim 140$ MeV. This indicates stronger suppression of the thermodynamic quantities in the strange sector which could be due to the (undiscovered) strange particles which 
are absent in the list considered. This situation can be remedied if we choose different (perhaps, smaller) values of $K_B$ for strange hadrons. However,  this will introduce additional parameter in the model which will render the MFHRG model more complex. It is important to note that HRG description breaks down near transition temperature $T\sim 155$ MeV and chiral  symmetry effects play an important role in the thermodynamics description of QCD matter\cite{Marczenko:2021icv,Koch:2023oez}.

%\begin{figure}[h]
%\vspace{-0.4cm}
%\hspace{-0.8cm}
%\begin{center}
%\begin{tabular}{c c}
 %\includegraphics[width=0.4\textwidth]{chiBQ11_chiB2_hrg.pdf}&
%  \includegraphics[width=0.4\textwidth]{chiBS11_chiB2_hrg.pdf}\\
 % \includegraphics[width=0.4\textwidth]{chiBQ11_chiB2.pdf}&
%\includegraphics[width=0.4\textwidth]{chiBS11_chiB2.pdf}
 %\end{tabular}
%\caption{Top panel  shows the ratios $\chi_{BQ}^{11}/\chi_{B}^2$ and $\chi_{BS}^{11}/\chi_{B}^2$ estimated within MFHRG model. In the bottom panel we compare HRG model estimates  with two different lists of hadrons, namely QMHRG2020 and HRG2016.  Lattice data has been taken from Ref.~\cite{Karthein:2021cmb}}.
	%	\label{bsqratio}
 % \end{center}
 %\end{figure}

 \begin{figure}[t]
 	\vspace{-0.4cm}
 	\begin{center}
 		\begin{tabular}{c c}
 			% \hspace {-0.7cm}
 			\includegraphics[scale=0.5]{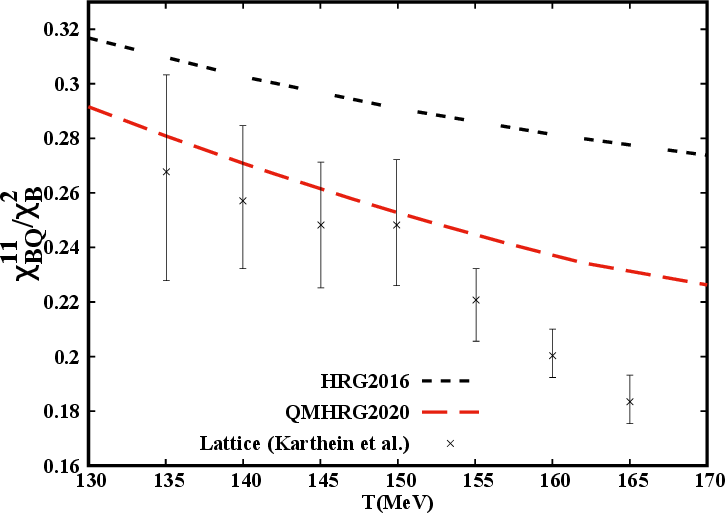}
 			\includegraphics[scale=0.5]{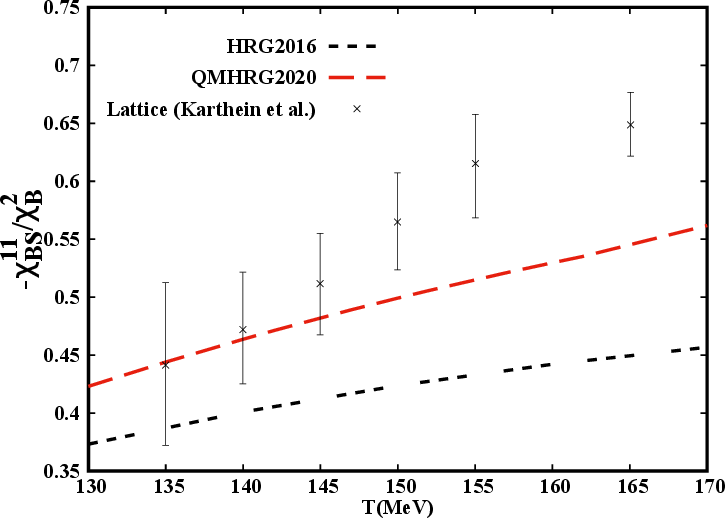} \\ 
 			%  \vspace{0.4cm}
 			\includegraphics[scale=0.5]{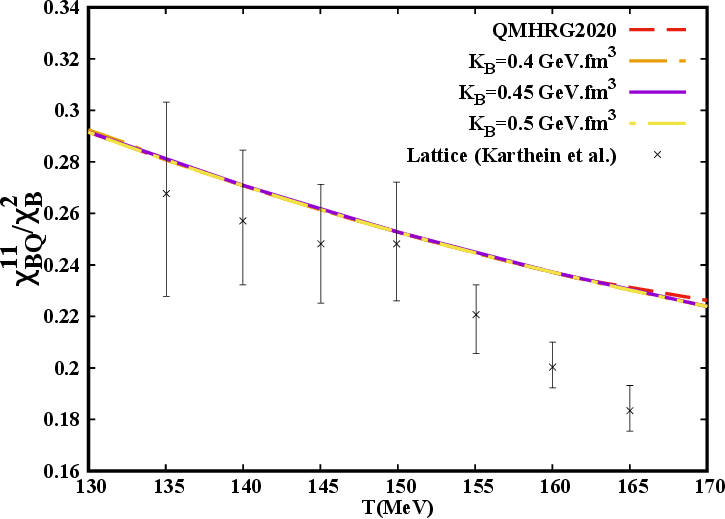}
 			\includegraphics[scale=0.5]{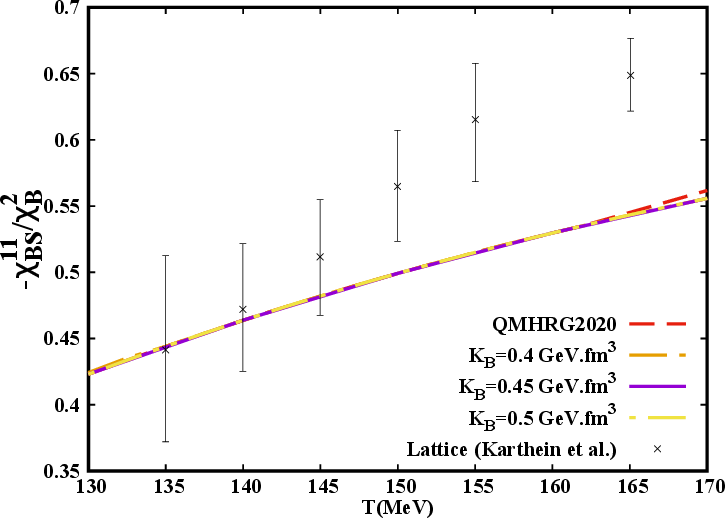} 
 		\end{tabular}
 	\caption{Top panel  shows the ratios $\chi_{BQ}^{11}/\chi_{B}^2$ and $\chi_{BS}^{11}/\chi_{B}^2$ estimated within MFHRG model. In the bottom panel we compare HRG model estimates  with two different lists of hadrons, namely QMHRG2020 and HRG2016.  Lattice data has been taken from Ref.~\cite{Karthein:2021cmb}}.
 	\label{bsqratio}
 	\end{center}
 \end{figure}

We next consider another interesting set of observable quantities, namely $\chi_{BQ}^{11}/\chi_B^2$ and $-\chi_{BS}^{11}/\chi_B^2$ as shown in Fig.\ref{bsqratio}.  Top panel shows the effect of additional resonances on these quantities. We get good agreement with the lattice data if we include all the known as well as unknown resonances in the partition function of HRG model. Bottom panel shows  $\chi_{BQ}^{11}/\chi_B^2$ and $-\chi_{BS}^{11}/\chi_B^2$ estimated within MFHRG model with varying strengths of repulsive interaction. We note that the effect of repulsive interaction accounted via mean-field interaction approximately cancels below $T_c$. This indicates that these ratios are independent of repulsive interactions and only the attractive interactions accounted through the inclusion of resonances in the partition function have any effect. Such cancellation has also been observed where the repulsive interactions are accounted via excluded volume formulation of HRG (EVHRG)~\cite{Karthein:2021cmb}. However, above $T_c$, a small deviation of these ratios estimated within MFHRG model from ideal HRG model can be noted. Reason for this deviation is that, these ratios depend on mean-field parameter only at very high temperatures.

\begin{figure}[t]
\vspace{-0.4cm}
\hspace{-0.8cm}
\begin{center}
\begin{tabular}{c c}
 \includegraphics[scale=0.5]{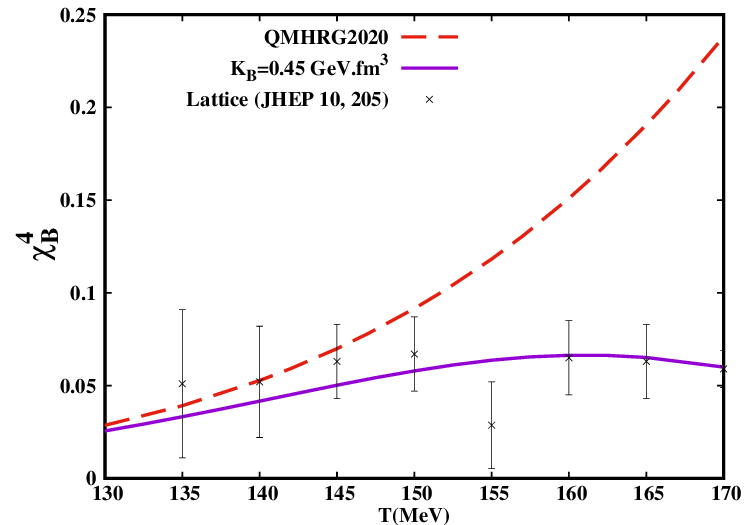} &
  \includegraphics[scale=0.5]{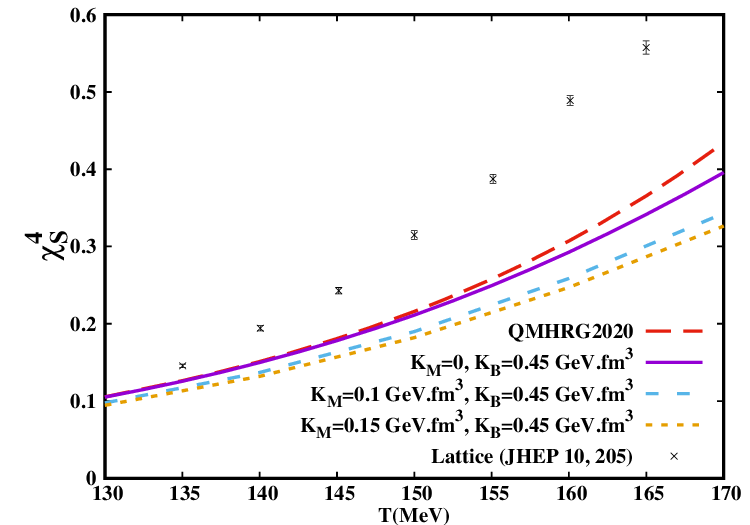}
  \end{tabular}
\caption{ Fourth order susceptibilities of conserved charge vs temperature at $\mu_B=\mu_S=0$. Lattice data has been taken from Ref.~\cite{Borsanyi:2018grb}}.
		\label{chi4}
  \end{center}
 \end{figure}

Fig.\ref{chi4} shows fourth order susceptibilities. We  note that MFHRG estimates of $\chi_B^4$ are in remarkable agreement with LQCD results over wide range of temperatures.  However,  HRG model and its extension underestimates $\chi_S^4$ which again indicates the necessity of unknown strange baryons.  We have not included $\chi_Q^4$ as reliable lattice data is not available for this quantity.

\begin{figure}[t]
\vspace{-0.4cm}
\hspace{-0.8cm}
\begin{center}
\begin{tabular}{c c}
 \includegraphics[scale=0.5]{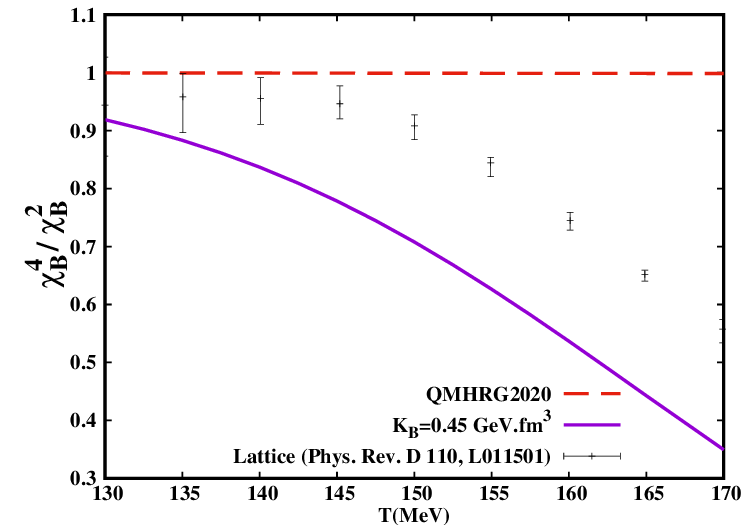}&
  \includegraphics[scale=0.5]{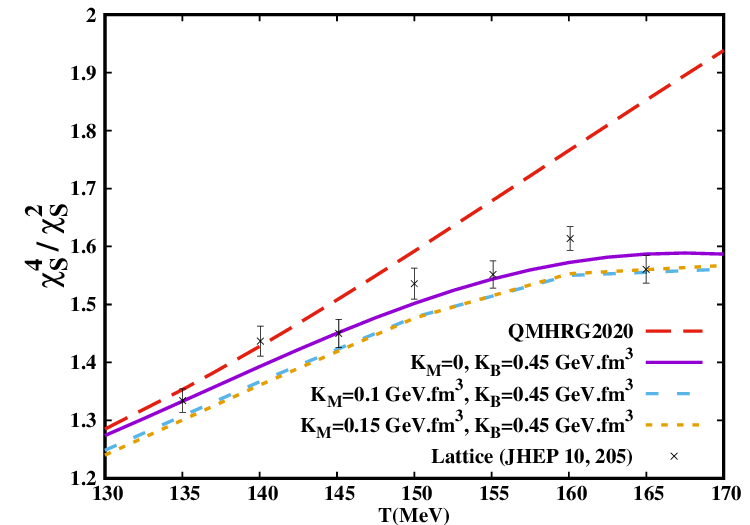}
  \end{tabular}
\caption{ Temperature dependence of ratios of second to fourth order susceptibilities of conserved charge  at $\mu_B=\mu_S=0$. LQCD data of $\chi_B^4/\chi_B^2$ is taken from the Ref.\cite{Borsanyi:2023wno} and that of $\chi_S^4/\chi_S^2$ is taken from the Ref. \cite{Borsanyi:2018grb}.}.
		\label{ratios}
  \end{center}
 \end{figure}

Fig.\ref{ratios} shows ratios of fourth and second order susceptibilities of various conserved charges. In particular, the ratio $\chi_B^4/\chi_B^2$ can be approximately given by

\begin{equation}
\frac{\chi_B^4}{\chi_B^2}\simeq 1-12 \frac{K_BT^2}{2}(\beta^3n_{B}^{\text{id}})
\label{chib_ratio}
\end{equation}

In case of ideal HRG this ratio is equal to one as shown by red dashed curve. However, presence of repulsive interaction reduces this ratio. In MFHRG model the deviation from HRG model results proportional to mean field parameter $K_B$ and we get good agreement with the LQCD results.

\begin{figure}[h]
\vspace{-0.4cm}
\hspace{-0.8cm}
\begin{center}
\begin{tabular}{c c}
 \includegraphics[scale=0.5]{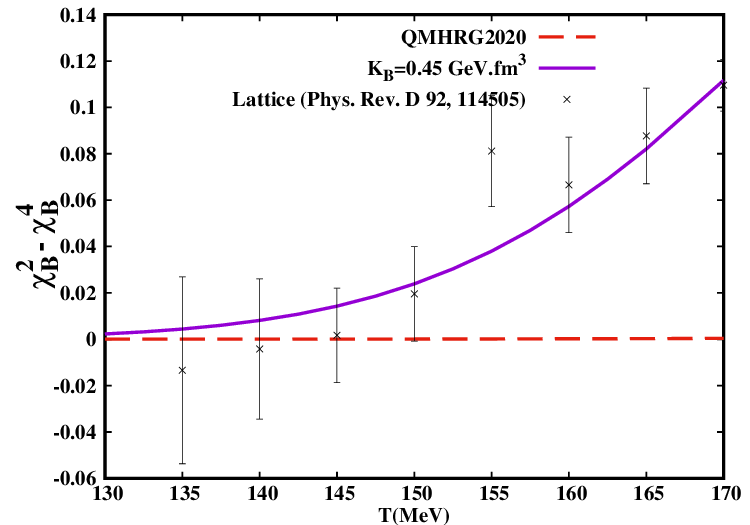}&
  \includegraphics[scale=0.5]{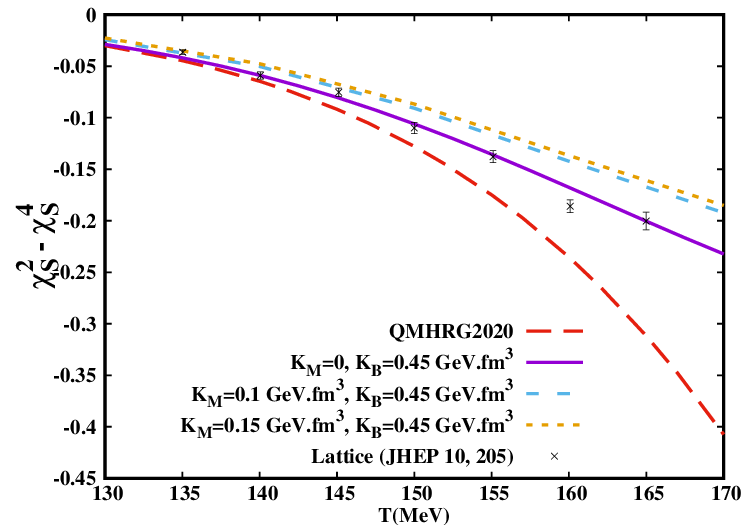}
  \end{tabular}
\caption{(Color online)Temperature dependence of difference of second to fourth order susceptibilities of conserved charge  at $\mu_B=\mu_S=0$. LQCD data of $\chi_B^2-\chi_B^4$ is taken from the Ref.\cite{Bellwied:2015lba} and that of $\chi_S^2-\chi_S^4$ is taken from the Ref. \cite{Borsanyi:2018grb}.}
		\label{diff}
  \end{center}
 \end{figure}

Separating the effects of repulsive interactions from other medium effects, namely in medium mass modifications and broadening of spectral width etc., is very difficult. It turns out that these effects are removed if we take differences of baryon susceptibilities\cite{Bazavov:2013dta}. Fig.\ref{diff} shows differences of second and fourth order charge susceptibilities. In case of baryon susceptibilities, it is easy to show that $\chi_B^2-\chi_B^4$ is approximately given by

\begin{equation}
\chi_B^2-\chi_B^4\simeq12 \frac{K_BT^2}{2}(\beta^3n_{B}^{\text{id}})
\label{chib_diff}
\end{equation}
 In the absence of repulsive interaction ($K_B=0$)  $\chi_B^2-\chi_B^4$ vanishes which is expected in  ideal HRG model. Effect of repulsive interaction is to increase this difference. MFHRG model with $K_B=0.45$ $\text{GeV.fm}^{3}$ reproduce LQCD results quite remarkably all the way up to $T=170$ MeV. MFHRG model is also in remarkable agreement with LQCD results of $\chi_S^2-\chi_S^4$ .  It may be noted that the differences in susceptibilities do not necessarily vanish in case of interacting theories; especially if the effects of chiral symmetry are included in the models describing hadronic phase\cite{Marczenko:2021icv,Koch:2023oez}.

 \begin{figure}[t]
 	\vspace{-0.4cm}
 	\begin{center}
 		\begin{tabular}{c c}
 			\includegraphics[scale=0.5]{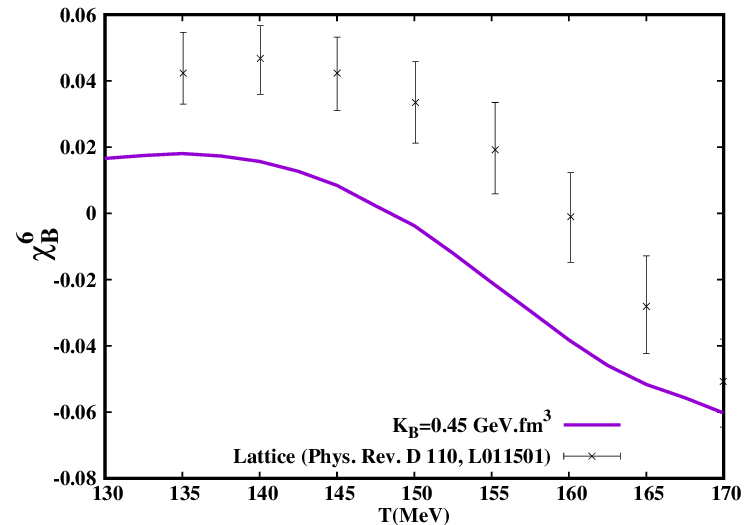}&
 			\includegraphics[scale=0.5]{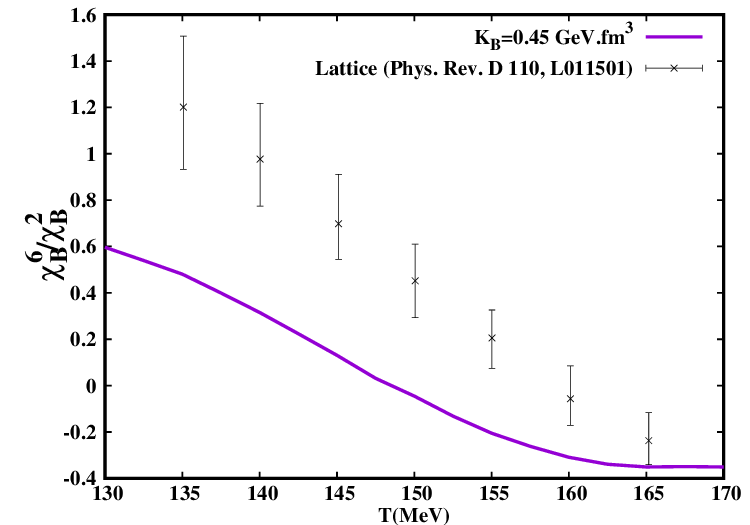}
 		\end{tabular}
 		\caption{(Color online)Temperature dependence of $\chi_B^6$ and  the ratio of sixth to second order susceptibilities  $\chi_B^6/\chi_B^2$. LQCD data is taken from the Ref.\cite{Borsanyi:2023wno}.}
 		\label{chiB6}
 	\end{center}
 \end{figure}
Fig.~\ref{chiB6} shows the variation of $\chi_B^6$ and $\chi_B^6/\chi_B^2$ as a function of temperature.  MFHRG model underestimate these quantities at low temperature; however the overall behavior is in good agreement with the lattice data.  It is to be noted that, higher order susceptibilities, are more sensitive to the modeling of the interactions among the particles. For instance, in case of parity doublet model\cite{Marczenko:2021icv}  which includes the chiral symmetry, the ratio $\chi_B^6/\chi_B^2$ shows strong sensitivity to dynamical effects related to the chiral symmetry restoration near $T_c$.

\section{Conclusion}
\label{sec4}

 We discussed the effect of repulsive interaction on the fluctuations and correlation of conserved charges, namely baryon number (B), electric charge (Q) and strangeness (S). We included the repulsive interaction in ideal HRG model via relativistic mean-field approach (MFHRG). In our study,  we considered a latest version of hadron list, namely QMHRG2020. We found that ideal HRG model is not sufficient to describe the LQCD data of conserved charge fluctuations and correlations especially near QCD transition temperature ($T_c$). We get good agreement with LQCD results if we include repulsive interactions whose role becomes more important near $T_c$. In our analysis we found that we have to consider a minimal extension of ideal HRG model in which we include the repulsive interactions (via mean-field approach) only for baryon-baryon and antibaryon-antibaryon pairs.  In this case, the repulsive interaction between meson-meson pairs and meson-baryon pairs is neglected. In fact,  interaction between meson-meson pairs and meson-baryon pairs is dominated by resonance formation which renders the attractive interactions already included in HRG model. However, a small but finite repulsion between meson-meson pairs is necessary to describe $\chi_Q^n$ and $\chi_{BQ}^{11}$. 
 
 Ratios of second to fourth order susceptibilities estimated within MFHRG model are also found to be in good agreement with LQCD results. Specifically, the ratio $\chi_{B}^{4}/\chi_{B}^2$ (Eq.\ref{chib_ratio}) deviates from its HRG model value ($=1$) as the temperature increases. This result is in agreement with the LQCD results especially near $T_c$.  We also found that, although the correlations $\chi_{BS}^{11}$ and $\chi_{BQ}^{11}$ depend on the repulsive interactions,  the ratios $\chi_{BS}^{11}/\chi_{B}^2$ and $\chi_{BQ}^{11}/\chi_B^2$ are independent of repulsive interactions. These later quantities only depend on the spectrum of hadrons included in the partition function of ideal HRG model. Further, in order to separate  the effects of repulsive interaction from other medium effects, we considered the differences in second and fourth order susceptibilities. Difference $\chi_B^2-\chi_B^4$ vanishes for ideal HRG and increases with temperature as we include the repulsive interaction (Eq.(\ref{chib_diff})) in agreement with the LQCD results.
 
 We finally conclude that the role of repulsive interaction in the description of QCD matter is very important especially near QCD transition temperature. The susceptibilities, correlations, their ratios and differences estimated within ambit of MFHRG are in good agreement with the LQCD results if we constrain mean-field parameter $K_B$ characterizing the repulsive interaction between baryon-baryon pairs and antibaryon-antibaryon pairs to $0.40\: \text{GeV.fm}^{3}\le K_B\le 0.450\: \text{GeV.fm}^{3}$. Mean-field parameter characterizing the repulsive interaction between mesons also play significant role in charge and strangeness susceptibilities. Small but non-zero value  of $K_M\sim 0.05$ $\text{GeV.fm}^{3}$ is required to get good agreement of MFHRG model with the lattice results for $\chi_Q^2$ and $\chi_S^2$. It is to be noted that, the susceptibilities and correlations related to strangeness show poor agreement with the lattice data, especially when repulsive interaction is incorporated. This underestimation of susceptibilities in the strangeness sector may be related to the absence of some strange hadrons (undiscovered) in the QMHRG2020 list. 
 
 \section{Acknowledgment}
 SP acknowledges financial support from the Department of Atomic Energy, India.
GK is financially supported by the DST-INSPIRE faculty award under  the grant number 
 DST/INSPIRE/04/2017/002293.

\bibliography{paper}

\begin{thebibliography}{10}

\bibitem{Ding:2015ona}
H.-T. Ding, F.~Karsch and S.~Mukherjee, {\em Int. J. Mod. Phys. E} {\bf 24},
  1530007  (2015), \href{http://arxiv.org/abs/1504.05274}{{\ttfamily
  arXiv:1504.05274 [hep-lat]}}.

\bibitem{Bzdak:2019pkr}
A.~Bzdak, S.~Esumi, V.~Koch, J.~Liao, M.~Stephanov and N.~Xu, {\em Phys. Rept.}
  {\bf 853}, 1  (2020), \href{http://arxiv.org/abs/1906.00936}{{\ttfamily
  arXiv:1906.00936 [nucl-th]}}.

\bibitem{Ratti:2021ubw}
C.~Ratti and R.~Bellwied, {\em {The Deconfinement Transition of QCD: Theory
  Meets Experiment}}, Lecture Notes in Physics, Vol.~981 6 2021.

\bibitem{Dashen:1969ep}
R.~Dashen, S.-K. Ma and H.~J. Bernstein, {\em Phys. Rev.} {\bf 187}, 345
  (1969).

\bibitem{Dashen:1974jw}
R.~F. Dashen and R.~Rajaraman, {\em Phys. Rev. D} {\bf 10},   694  (1974).

\bibitem{Welke:1990za}
G.~M. Welke, R.~Venugopalan and M.~Prakash, {\em Phys. Lett. B} {\bf 245}, 137
  (1990).

\bibitem{Venugopalan:1992hy}
R.~Venugopalan and M.~Prakash, {\em Nucl. Phys. A} {\bf 546}, 718  (1992).

\bibitem{braun2004quark}
P.~Braun-Munzinger, K.~Redlich and J.~Stachel, Quark gluon plasma 3  (2004).

\bibitem{Bhattacharyya:2015pra}
A.~Bhattacharyya, S.~K. Ghosh, R.~Ray and S.~Samanta, {\em EPL} {\bf 115},
  62003  (2016), \href{http://arxiv.org/abs/1504.04533}{{\ttfamily
  arXiv:1504.04533 [hep-ph]}}.

\bibitem{Kadam:2019rzo}
G.~Kadam, S.~Pal and A.~Bhattacharyya, {\em J. Phys. G} {\bf 47},   125106
  (2020), \href{http://arxiv.org/abs/1908.10618}{{\ttfamily arXiv:1908.10618
  [hep-ph]}}.

\bibitem{Pradhan:2021vtp}
G.~S. Pradhan, D.~Sahu, S.~Deb and R.~Sahoo, {\em J. Phys. G} {\bf 50},
  055104  (2023), \href{http://arxiv.org/abs/2106.14297}{{\ttfamily
  arXiv:2106.14297 [hep-ph]}}.

\bibitem{Braun-Munzinger:1994ewq}
P.~Braun-Munzinger, J.~Stachel, J.~P. Wessels and N.~Xu, {\em Phys. Lett. B}
  {\bf 344}, 43  (1995), \href{http://arxiv.org/abs/nucl-th/9410026}{{\ttfamily
  arXiv:nucl-th/9410026}}.

\bibitem{Braun-Munzinger:1995uec}
P.~Braun-Munzinger, J.~Stachel, J.~P. Wessels and N.~Xu, {\em Phys. Lett. B}
  {\bf 365}, 1  (1996), \href{http://arxiv.org/abs/nucl-th/9508020}{{\ttfamily
  arXiv:nucl-th/9508020}}.

\bibitem{Braun-Munzinger:1999hun}
P.~Braun-Munzinger, I.~Heppe and J.~Stachel, {\em Phys. Lett. B} {\bf 465}, 15
  (1999), \href{http://arxiv.org/abs/nucl-th/9903010}{{\ttfamily
  arXiv:nucl-th/9903010}}.

\bibitem{Cleymans:1996cd}
J.~Cleymans, D.~Elliott, H.~Satz and R.~L. Thews, {\em Z. Phys. C} {\bf 74},
  319  (1997), \href{http://arxiv.org/abs/nucl-th/9603004}{{\ttfamily
  arXiv:nucl-th/9603004}}.

\bibitem{Cleymans:1999st}
J.~Cleymans and K.~Redlich, {\em Phys. Rev. C} {\bf 60},   054908  (1999),
  \href{http://arxiv.org/abs/nucl-th/9903063}{{\ttfamily
  arXiv:nucl-th/9903063}}.

\bibitem{Becattini:2005xt}
F.~Becattini, J.~Manninen and M.~Gazdzicki, {\em Phys. Rev. C} {\bf 73},
  044905  (2006), \href{http://arxiv.org/abs/hep-ph/0511092}{{\ttfamily
  arXiv:hep-ph/0511092}}.

\bibitem{Braun-Munzinger:2001hwo}
P.~Braun-Munzinger, D.~Magestro, K.~Redlich and J.~Stachel, {\em Phys. Lett. B}
  {\bf 518}, 41  (2001), \href{http://arxiv.org/abs/hep-ph/0105229}{{\ttfamily
  arXiv:hep-ph/0105229}}.

\bibitem{Andronic:2005yp}
A.~Andronic, P.~Braun-Munzinger and J.~Stachel, {\em Nucl. Phys. A} {\bf 772},
  167  (2006), \href{http://arxiv.org/abs/nucl-th/0511071}{{\ttfamily
  arXiv:nucl-th/0511071}}.

\bibitem{Andronic:2008gu}
A.~Andronic, P.~Braun-Munzinger and J.~Stachel, {\em Phys. Lett. B} {\bf 673},
  142  (2009), \href{http://arxiv.org/abs/0812.1186}{{\ttfamily arXiv:0812.1186
  [nucl-th]}}, [Erratum: Phys.Lett.B 678, 516 (2009)].

\bibitem{Karsch:2003vd}
F.~Karsch, K.~Redlich and A.~Tawfik, {\em Eur. Phys. J. C} {\bf 29}, 549
  (2003), \href{http://arxiv.org/abs/hep-ph/0303108}{{\ttfamily
  arXiv:hep-ph/0303108}}.

\bibitem{Karsch:2003zq}
F.~Karsch, K.~Redlich and A.~Tawfik, {\em Phys. Lett. B} {\bf 571}, 67  (2003),
  \href{http://arxiv.org/abs/hep-ph/0306208}{{\ttfamily arXiv:hep-ph/0306208}}.

\bibitem{Tawfik:2004sw}
A.~Tawfik, {\em Phys. Rev. D} {\bf 71},   054502  (2005),
  \href{http://arxiv.org/abs/hep-ph/0412336}{{\ttfamily arXiv:hep-ph/0412336}}.

\bibitem{Huovinen:2009yb}
P.~Huovinen and P.~Petreczky, {\em Nucl. Phys. A} {\bf 837}, 26  (2010),
  \href{http://arxiv.org/abs/0912.2541}{{\ttfamily arXiv:0912.2541 [hep-ph]}}.

\bibitem{Alba:2015iva}
P.~Alba, R.~Bellwied, M.~Bluhm, V.~Mantovani~Sarti, M.~Nahrgang and C.~Ratti,
  {\em Phys. Rev. C} {\bf 92},   064910  (2015),
  \href{http://arxiv.org/abs/1504.03262}{{\ttfamily arXiv:1504.03262
  [hep-ph]}}.

\bibitem{Huovinen:2017ogf}
P.~Huovinen and P.~Petreczky, {\em Phys. Lett. B} {\bf 777}, 125  (2018),
  \href{http://arxiv.org/abs/1708.00879}{{\ttfamily arXiv:1708.00879
  [hep-ph]}}.

\bibitem{Vovchenko:2014pka}
V.~Vovchenko, D.~V. Anchishkin and M.~I. Gorenstein, {\em Phys. Rev. C} {\bf
  91},   024905  (2015), \href{http://arxiv.org/abs/1412.5478}{{\ttfamily
  arXiv:1412.5478 [nucl-th]}}.

\bibitem{Bazavov:2013uja}
A.~Bazavov, H.~T. Ding, P.~Hegde, F.~Karsch, C.~Miao, S.~Mukherjee,
  P.~Petreczky, C.~Schmidt and A.~Velytsky, {\em Phys. Rev. D} {\bf 88},
  094021  (2013), \href{http://arxiv.org/abs/1309.2317}{{\ttfamily
  arXiv:1309.2317 [hep-lat]}}.

\bibitem{Bhattacharyya:2013oya}
A.~Bhattacharyya, S.~Das, S.~K. Ghosh, R.~Ray and S.~Samanta, {\em Phys. Rev.
  C} {\bf 90},   034909  (2014),
  \href{http://arxiv.org/abs/1310.2793}{{\ttfamily arXiv:1310.2793 [hep-ph]}}.

\bibitem{Bhattacharyya:2017gwt}
A.~Bhattacharyya, S.~K. Ghosh, S.~Maity, S.~Raha, R.~Ray, K.~Saha, S.~Samanta
  and S.~Upadhaya, {\em Phys. Rev. C} {\bf 99},   045207  (2019),
  \href{http://arxiv.org/abs/1708.04549}{{\ttfamily arXiv:1708.04549
  [hep-ph]}}.

\bibitem{Bhattacharyya:2014uxa}
A.~Bhattacharyya, R.~Ray and S.~Sur, {\em Phys. Rev. D} {\bf 91},   051501
  (2015), \href{http://arxiv.org/abs/1412.8316}{{\ttfamily arXiv:1412.8316
  [hep-ph]}}.

\bibitem{Fukushima:2008wg}
K.~Fukushima, {\em Phys. Rev. D} {\bf 77},   114028  (2008),
  \href{http://arxiv.org/abs/0803.3318}{{\ttfamily arXiv:0803.3318 [hep-ph]}},
  [Erratum: Phys.Rev.D 78, 039902 (2008)].

\bibitem{Roessner:2006xn}
S.~Roessner, C.~Ratti and W.~Weise, {\em Phys. Rev. D} {\bf 75},   034007
  (2007), \href{http://arxiv.org/abs/hep-ph/0609281}{{\ttfamily
  arXiv:hep-ph/0609281}}.

\bibitem{Sasaki:2006ww}
C.~Sasaki, B.~Friman and K.~Redlich, {\em Phys. Rev. D} {\bf 75},   074013
  (2007), \href{http://arxiv.org/abs/hep-ph/0611147}{{\ttfamily
  arXiv:hep-ph/0611147}}.

\bibitem{Ratti:2007jf}
C.~Ratti, S.~Roessner and W.~Weise, {\em Phys. Lett. B} {\bf 649}, 57  (2007),
  \href{http://arxiv.org/abs/hep-ph/0701091}{{\ttfamily arXiv:hep-ph/0701091}}.

\bibitem{Fukushima:2009dx}
K.~Fukushima, {\em Phys. Rev. D} {\bf 79},   074015  (2009),
  \href{http://arxiv.org/abs/0901.0783}{{\ttfamily arXiv:0901.0783 [hep-ph]}}.

\bibitem{Schaefer:2006ds}
B.-J. Schaefer and J.~Wambach, {\em Phys. Rev. D} {\bf 75},   085015  (2007),
  \href{http://arxiv.org/abs/hep-ph/0603256}{{\ttfamily arXiv:hep-ph/0603256}}.

\bibitem{Schaefer:2009ui}
B.-J. Schaefer, M.~Wagner and J.~Wambach, {\em Phys. Rev. D} {\bf 81},   074013
   (2010), \href{http://arxiv.org/abs/0910.5628}{{\ttfamily arXiv:0910.5628
  [hep-ph]}}.

\bibitem{Schaefer:2007pw}
B.-J. Schaefer, J.~M. Pawlowski and J.~Wambach, {\em Phys. Rev. D} {\bf 76},
  074023  (2007), \href{http://arxiv.org/abs/0704.3234}{{\ttfamily
  arXiv:0704.3234 [hep-ph]}}.

\bibitem{Wambach:2009ee}
J.~Wambach, B.-J. Schaefer and M.~Wagner, {\em Acta Phys. Polon. Supp.} {\bf
  3}, 691  (2010), \href{http://arxiv.org/abs/0911.0296}{{\ttfamily
  arXiv:0911.0296 [hep-ph]}}.

\bibitem{Motornenko:2020yme}
A.~Motornenko, S.~Pal, A.~Bhattacharyya, J.~Steinheimer and H.~Stoecker, {\em
  Phys. Rev. C} {\bf 103},   054908  (2021),
  \href{http://arxiv.org/abs/2009.10848}{{\ttfamily arXiv:2009.10848
  [hep-ph]}}.

\bibitem{Pal:2020ink}
S.~Pal, A.~Bhattacharyya and R.~Ray, {\em Nucl. Phys. A} {\bf 1010},   122177
  (2021), \href{http://arxiv.org/abs/2006.08985}{{\ttfamily arXiv:2006.08985
  [hep-ph]}}.

\bibitem{Pal:2020ucy}
S.~Pal, G.~Kadam, H.~Mishra and A.~Bhattacharyya, {\em Phys. Rev. D} {\bf 103},
    054015  (2021), \href{http://arxiv.org/abs/2010.10761}{{\ttfamily
  arXiv:2010.10761 [hep-ph]}}.

\bibitem{ab1}
A.~Bhattacharyya, S.~K. Ghosh, A.~Lahiri, S.~Majumder, S.~Raha and R.~Ray, {\em
  Phys. Rev. C} {\bf 89},   064905  (2014),
  \href{http://arxiv.org/abs/1212.6134}{{\ttfamily arXiv:1212.6134 [hep-ph]}}.

\bibitem{ab2}
A.~Bhattacharyya, S.~K. Ghosh, R.~Ray, K.~Saha and S.~Upadhyay, {\em EPL} {\bf
  116},   52001  (2016), \href{http://arxiv.org/abs/1507.08795}{{\ttfamily
  arXiv:1507.08795 [hep-ph]}}.

\bibitem{beta}
A.~Bhattacharyya, S.~K. Ghosh, S.~Majumder and R.~Ray, {\em Phys. Rev. D} {\bf
  86},   096006  (2012), \href{http://arxiv.org/abs/1107.5941}{{\ttfamily
  arXiv:1107.5941 [hep-ph]}}.

\bibitem{Pal2024}
S.~Pal, A.~Motornenko, V.~Vovchenko, A.~Bhattacharyya, J.~Steinheimer and
  H.~Stoecker, {\em Phys. Rev. D} {\bf 109},   014009  (2024),
  \href{http://arxiv.org/abs/2306.10596}{{\ttfamily arXiv:2306.10596
  [hep-ph]}}.

\bibitem{PhysRevC.96.025205}
A.~Mukherjee, J.~Steinheimer and S.~Schramm, {\em Phys. Rev. C} {\bf 96},
  025205 (Aug 2017).

\bibitem{RAU2014176}
P.~Rau, J.~Steinheimer, S.~Schramm and H.~Stöcker, {\em Physics Letters B}
  {\bf 733}, 176  (2014).

\bibitem{PhysRevD.97.114030}
V.~Vovchenko, J.~Steinheimer, O.~Philipsen and H.~Stoecker, {\em Phys. Rev. D}
  {\bf 97},   114030 (Jun 2018).

\bibitem{Petreczky:2012rq}
P.~Petreczky, {\em J. Phys. G} {\bf 39},   093002  (2012),
  \href{http://arxiv.org/abs/1203.5320}{{\ttfamily arXiv:1203.5320 [hep-lat]}}.

\bibitem{Borsanyi:2010bp}
Wuppertal-Budapest Collaboration, S.~Borsanyi, Z.~Fodor, C.~Hoelbling, S.~D.
  Katz, S.~Krieg, C.~Ratti and K.~K. Szabo, {\em JHEP} {\bf 09},   073  (2010),
  \href{http://arxiv.org/abs/1005.3508}{{\ttfamily arXiv:1005.3508 [hep-lat]}}.

\bibitem{Borsanyi:2011sw}
S.~Borsanyi, Z.~Fodor, S.~D. Katz, S.~Krieg, C.~Ratti and K.~Szabo, {\em JHEP}
  {\bf 01},   138  (2012), \href{http://arxiv.org/abs/1112.4416}{{\ttfamily
  arXiv:1112.4416 [hep-lat]}}.

\bibitem{HotQCD:2012fhj}
HotQCD Collaboration, A.~Bazavov {\em et~al.}, {\em Phys. Rev. D} {\bf 86},
  034509  (2012), \href{http://arxiv.org/abs/1203.0784}{{\ttfamily
  arXiv:1203.0784 [hep-lat]}}.

\bibitem{Bazavov:2013dta}
A.~Bazavov {\em et~al.}, {\em Phys. Rev. Lett.} {\bf 111},   082301  (2013),
  \href{http://arxiv.org/abs/1304.7220}{{\ttfamily arXiv:1304.7220 [hep-lat]}}.

\bibitem{Bazavov:2014xya}
A.~Bazavov {\em et~al.}, {\em Phys. Rev. Lett.} {\bf 113},   072001  (2014),
  \href{http://arxiv.org/abs/1404.6511}{{\ttfamily arXiv:1404.6511 [hep-lat]}}.

\bibitem{Bazavov:2014yba}
A.~Bazavov {\em et~al.}, {\em Phys. Lett. B} {\bf 737}, 210  (2014),
  \href{http://arxiv.org/abs/1404.4043}{{\ttfamily arXiv:1404.4043 [hep-lat]}}.

\bibitem{Borsanyi:2014ewa}
S.~Borsanyi, Z.~Fodor, S.~D. Katz, S.~Krieg, C.~Ratti and K.~K. Szabo, {\em
  Phys. Rev. Lett.} {\bf 113},   052301  (2014),
  \href{http://arxiv.org/abs/1403.4576}{{\ttfamily arXiv:1403.4576 [hep-lat]}}.

\bibitem{Bellwied:2015lba}
R.~Bellwied, S.~Borsanyi, Z.~Fodor, S.~D. Katz, A.~Pasztor, C.~Ratti and K.~K.
  Szabo, {\em Phys. Rev. D} {\bf 92},   114505  (2015),
  \href{http://arxiv.org/abs/1507.04627}{{\ttfamily arXiv:1507.04627
  [hep-lat]}}.

\bibitem{Ding:2015fca}
H.~T. Ding, S.~Mukherjee, H.~Ohno, P.~Petreczky and H.~P. Schadler, {\em Phys.
  Rev. D} {\bf 92},   074043  (2015),
  \href{http://arxiv.org/abs/1507.06637}{{\ttfamily arXiv:1507.06637
  [hep-lat]}}.

\bibitem{Bazavov:2015zja}
A.~Bazavov {\em et~al.}, {\em Phys. Rev. D} {\bf 93},   014512  (2016),
  \href{http://arxiv.org/abs/1509.05786}{{\ttfamily arXiv:1509.05786
  [hep-lat]}}.

\bibitem{DElia:2016jqh}
M.~D'Elia, G.~Gagliardi and F.~Sanfilippo, {\em Phys. Rev. D} {\bf 95},
  094503  (2017), \href{http://arxiv.org/abs/1611.08285}{{\ttfamily
  arXiv:1611.08285 [hep-lat]}}.

\bibitem{Bazavov:2017dus}
A.~Bazavov {\em et~al.}, {\em Phys. Rev. D} {\bf 95},   054504  (2017),
  \href{http://arxiv.org/abs/1701.04325}{{\ttfamily arXiv:1701.04325
  [hep-lat]}}.

\bibitem{Borsanyi:2018grb}
S.~Borsanyi, Z.~Fodor, J.~N. Guenther, S.~K. Katz, K.~K. Szabo, A.~Pasztor,
  I.~Portillo and C.~Ratti, {\em JHEP} {\bf 10},   205  (2018),
  \href{http://arxiv.org/abs/1805.04445}{{\ttfamily arXiv:1805.04445
  [hep-lat]}}.

\bibitem{Karthein:2021cmb}
J.~M. Karthein, V.~Koch, C.~Ratti and V.~Vovchenko, {\em Phys. Rev. D} {\bf
  104},   094009  (2021), \href{http://arxiv.org/abs/2107.00588}{{\ttfamily
  arXiv:2107.00588 [nucl-th]}}.

\bibitem{Rischke:1992rk}
D.~H. Rischke, J.~Schaffner, M.~I. Gorenstein, A.~Schaefer, H.~Stoecker and
  W.~Greiner, {\em Z. Phys. C} {\bf 56}, 325  (1992).

\bibitem{Vovchenko:2020lju}
V.~Vovchenko, {\em Int. J. Mod. Phys. E} {\bf 29},   2040002  (2020),
  \href{http://arxiv.org/abs/2004.06331}{{\ttfamily arXiv:2004.06331
  [nucl-th]}}.

\bibitem{Kapusta:1982qd}
J.~I. Kapusta and K.~A. Olive, {\em Nucl. Phys. A} {\bf 408}, 478  (1983).

\bibitem{Olive:1980dy}
K.~A. Olive, {\em Nucl. Phys. B} {\bf 190}, 483  (1981).

\bibitem{Anchishkin:2014hfa}
D.~Anchishkin and V.~Vovchenko, {\em J. Phys. G} {\bf 42},   105102  (2015),
  \href{http://arxiv.org/abs/1411.1444}{{\ttfamily arXiv:1411.1444 [nucl-th]}}.

\bibitem{Bollweg:2021vqf}
HotQCD Collaboration, D.~Bollweg, J.~Goswami, O.~Kaczmarek, F.~Karsch,
  S.~Mukherjee, P.~Petreczky, C.~Schmidt and P.~Scior, {\em Phys. Rev. D} {\bf
  104},  074512(2021), \href{http://arxiv.org/abs/2107.10011}{{\ttfamily
  arXiv:2107.10011 [hep-lat]}}.

\bibitem{Savchuk:2020yxc}
O.~Savchuk, Y.~Bondar, O.~Stashko, R.~V. Poberezhnyuk, V.~Vovchenko, M.~I.
  Gorenstein and H.~Stoecker, {\em Phys. Rev. C} {\bf 102},   035202  (2020),
  \href{http://arxiv.org/abs/2004.09004}{{\ttfamily arXiv:2004.09004
  [hep-ph]}}.

\bibitem{Andronic:2012ut}
A.~Andronic, P.~Braun-Munzinger, J.~Stachel and M.~Winn, {\em Phys. Lett. B}
  {\bf 718}, 80  (2012), \href{http://arxiv.org/abs/1201.0693}{{\ttfamily
  arXiv:1201.0693 [nucl-th]}}.

\bibitem{Sollfrank:1996hd}
J.~Sollfrank, P.~Huovinen, M.~Kataja, P.~V. Ruuskanen, M.~Prakash and
  R.~Venugopalan, {\em Phys. Rev. C} {\bf 55}, 392  (1997),
  \href{http://arxiv.org/abs/nucl-th/9607029}{{\ttfamily
  arXiv:nucl-th/9607029}}.

\bibitem{Vovchenko:2016rkn}
V.~Vovchenko, M.~I. Gorenstein and H.~Stoecker, {\em Phys. Rev. Lett.} {\bf
  118},   182301  (2017), \href{http://arxiv.org/abs/1609.03975}{{\ttfamily
  arXiv:1609.03975 [hep-ph]}}.

\bibitem{Marczenko:2021icv}
M.~Marczenko, K.~Redlich and C.~Sasaki, {\em Phys. Rev. D} {\bf 103},   054035
  (2021).

\bibitem{Koch:2023oez}
V.~Koch, M.~Marczenko, K.~Redlich and C.~Sasaki, {\em Phys. Rev. D} {\bf 109},
   014033  (2024), \href{http://arxiv.org/abs/2308.15794}{{\ttfamily
  arXiv:2308.15794 [hep-ph]}}.

\bibitem{Borsanyi:2023wno}
S.~Borsanyi, Z.~Fodor, J.~N. Guenther, S.~D. Katz, P.~Parotto, A.~Pasztor,
  D.~Pesznyak, K.~K. Szabo and C.~H. Wong, {\em Phys. Rev. D} {\bf 110},
  L011501  (2024), \href{http://arxiv.org/abs/2312.07528}{{\ttfamily
  arXiv:2312.07528 [hep-lat]}}.

\end{thebibliography}

\end{document}